\documentstyle[11pt,epsf]{article}
\newcommand{\postscript}[2]
   {\setlength{\epsfxsize}{#2\hsize}
   \centerline{\epsfbox{#1}}}
\begin{document}
\def\theequation {\thesection.\arabic{equation}}
\makeatletter\@addtoreset {equation}{section}\makeatother
\title{\bf Continuous families of embedded solitons in the third-order nonlinear
Schr\"odinger equation}
\author{Jianke Yang$^\dagger$, \hspace{0.2cm} 
T.R. Akylas$^{\dagger\dagger}$\\ \\
{\small $^\dagger$ Department of Mathematics and Statistics, University of Vermont, 
Burlington, VT 05401, USA} \\ \\
{\small $^{\dagger\dagger}$ Department of Mechanical Engineering, MIT, Cambridge, MA 02139, USA
}}
\date{\empty}
\maketitle
\thispagestyle{empty}

\begin{abstract}
The nonlinear Schr\"odinger equation with a third-order dispersive term is considered. 
Infinite families of embedded solitons, parameterized by the propagation velocity,
are found through a gauge transformation. 
By applying this transformation, an embedded soliton can acquire any 
velocity above a certain threshold value. 
It is also shown that all these families of embedded solitons are linearly stable, 
but nonlinearly semi-stable.

\end{abstract}


\section{Introduction}
The nonlinear Schr\"odinger (NLS) equation with a third-order dispersive term 
\begin{equation} \label{phiold}
i\phi_t+\phi_{xx}+|\phi|^2\phi=i\beta \phi_{xxx}
\end{equation}
arises in a wide variety of physical systems. 
For instance, propagation of pico-second optical pulses near the
zero second-order dispersion point
in an optical fiber is governed by this equation \cite{agrawal,hasegawa,wai90}. 
Femto-second pulses in a fiber laser cavity are modeled by this equation as well 
\cite{agrawal,hasegawa}. 
Eq. (\ref{phiold}) also arises in water waves near a caustic \cite{akylaskung}. 
With the rescaling of variables 
$\phi'=\beta \phi, x'=x/\beta, t'=t/\beta^2$, 
and dropping the primes, Eq. (\ref{phiold}) is normalized as 
\begin{equation} \label{phi}
i\phi_t+\phi_{xx}+|\phi|^2\phi=i\phi_{xxx}. 
\end{equation}
This is the third-order nonlinear Schr\"odinger (TNLS) equation we will study in this paper. 
Note that this equation is Hamiltonian. 

Solitary waves of the TNLS equation 
and their stability properties raise important issues from both 
physical and mathematical points of view. Physically, if stable solitary waves
exist in this equation, they could be used as information bits in communication systems
near the zero dispersion wavelength, where the fiber's second-order dispersion is 
relatively small. 
Mathematically, due to the third-order dispersive term, any solitary wave 
of the TNLS equation, 
if it exists, resides inside the continuous spectrum of that equation, and is thus an 
embedded soliton \cite{YangPRL99}. Classification of embedded solitons in the TNLS equation
and characterization of their stability properties pose non-trivial
mathematical challenges. Some progress has been made on these problems. 
For instance, Jang \& Benney \cite{jang}, 
Wai, et al. \cite{wai90} and Haus, et al. \cite{hausTOD} have shown 
that in the presence of third-order dispersion, the familiar NLS soliton with
the ``sech'' profile 
cannot remain stationary and will always shed continuous-wave radiation and lose energy. 
Akylas and Kung \cite{akylaskung}, Klauder, et al. \cite{klauder} and Calvo and Akylas 
\cite{calvoakylas1,calvoakylas2}) have discovered an infinite number of isolated 
embedded solitons with multi-hump profiles. 
With regard to stability, 
the numerical works by Klauder, et al. \cite{klauder} and Calvo \& Akylas \cite{calvoakylas2} 
have suggested that embedded solitons are linearly unstable, but nonlinearity 
has a stabilizing effect. 
As discussed below, however, those numerical results invite
a different interpretation from what these authors proposed. 

Despite the above progress, several important questions still remain open: 
are embedded solitons in the TNLS equation isolated, or they exist as continuous
families? are embedded solitons indeed linearly unstable as previously claimed
\cite{klauder,calvoakylas2}? are these solitons nonlinearly stable? 

From a broader perspective, there are many recent results in the literature
that bear upon these questions. 
For instance, embedded solitons have been discovered
in various physical systems such as
the fifth-order Korteweg--de Vries (KdV)
equations \cite{calvoakylas1,olver,groves,YangStudies,tan}, the extended nonlinear 
Schr\"odinger equations \cite{buryak,fujioka}, the coupled KdV
equations \cite{grimshaw}, the second-harmonic-generation (SHG)
system \cite{YangPRL99,PeliYang}, the massive Thirring model 
\cite{thirring,movingES}, the three-wave system \cite{threewave}, 
and others \cite{ChampneysYang,YangChampneys}. 
A common feature of all those embedded solitons is that they exist
at isolated parameter points. 
For such isolated embedded solitons, heuristic arguments \cite{YangPRL99} as well as 
rigorous soliton-perturbation \cite{YangStudies} and internal-perturbation 
\cite{tan,PeliYang} calculations have shown that they 
are always nonlinearly semi-stable if they are linearly stable. 
These analytical calculations are fully supported by direct numerical simulations
\cite{YangPRL99,tan,PeliYang,ChampneysYang,YangChampneys}. 
An outstanding open question, however, 
is whether embedded solitons can exist as continuous families. 
If they do, that would have important implications for their nonlinear stability properties: 
first, the previous arguments for semi-stability of isolated embedded solitons 
no longer apply. Second, when continuous embedded solitons are perturbed, 
they may, in principle, shed some energy via radiation and approach nearby embedded solitons. 
Thus, there is a possibility that continuous embedded solitons may be nonlinearly stable ---
a result which would be very significant physically. 
So far, questions of whether 
continuous families of embedded solitons are possible in physical systems 
and their stability properties
have not received much attention in the literature. 
Here, these issues will be discussed in the context of the
TNLS equation (\ref{phi}). 

In the present article, we will first show that 
infinite families of embedded solitons can be found 
in the TNLS equation through a gauge transformation. 
By applying this transformation, an embedded soliton can acquire any velocity above
a certain threshold value. 
Moreover, all these families of embedded solitons are linearly stable, 
contrary to previous claims about their linear instability. 
Lastly, we will numerically establish that these embedded solitons are still 
nonlinearly semi-stable. In other words, even though these solitons 
exist as continuous families, they still suffer nonlinear instability for
certain types of perturbations. 

\section{Infinite families of embedded solitons}
In this section, we study solitary waves of the TNLS equation (\ref{phi}). 
It is easy to see that due to the third-order dispersive term, 
any solitary wave of the TNLS equation is embedded in the continuous spectrum
of the linear part of that equation, thus is an embedded soliton \cite{YangPRL99}. 
We look for embedded solitons of the form
$\phi(x,t)=\psi (x-vt)e^{ikx+i\lambda t}$, 
where $\psi$ is a complex function, while $v, k$ and $\lambda$
are velocity, wavenumber and frequency constants. This furnishes moving 
embedded solitons of a rather general form, 
as more complicated phase and
amplitude functions generally lead to non-stationary propagation of the wave 
amplitude profile. For instance, inclusion of a chirp ($ix^2$ term) in the phase
function induces amplitude breathing \cite{uedakath}. 
With a redefinition of the function $\psi$ and frequency $\lambda$, 
the above soliton form can be rewritten as 
\begin{equation} \label{phiform}
\phi(x,t)=\psi(\theta) e^{i\lambda t}, 
\end{equation}
where 
\begin{equation} \label{theta}
\theta=x-vt. 
\end{equation}
In other words, the wavenumber $k$ in the original soliton form can be normalized to zero. 
Then $\psi(\theta)$ satisfies the equation
\begin{equation} \label{psi}
\psi_{\theta\theta}-\lambda \psi+|\psi|^2\psi=i(\psi_{\theta\theta\theta}+v\psi_\theta).
\end{equation}
It is noted that if $\psi(\theta)$ is a solution, so are $\psi(\theta)e^{i\delta}$, 
$\psi(\theta-\theta_0)$, and $\psi^*(-\theta)$. Here $\delta$ and $\theta_0$ are
arbitrary constants, and $\psi^*$ is the complex conjugate of $\psi$. 
Thus, without loss of generality, we require the solution $\psi$ to possess the
following symmetry: 
\begin{equation} \label{symmetry}
\psi(-\theta)=\psi^*(\theta), 
\end{equation}
i.e., Re($\psi$) is symmetric, and Im($\psi$) anti-symmetric. Numerically, 
$\psi$ can be determined by a shooting procedure with 
the following boundary conditions imposed: 
\begin{equation} \label{psiinf}
\psi(\theta) \longrightarrow 0, \;\;\; |\theta| \to \infty, 
\end{equation}
\begin{equation} \label{psizero}
\mbox{Re}(\psi_\theta)=\mbox{Im}(\psi)=\mbox{Im}(\psi_{\theta\theta})=0, \;\;\; \theta=0.
\end{equation}

Embedded solitons in Eq. (\ref{psi}) depend on two real parameters: 
the velocity $v$, and the frequency $\lambda$. 
We will show that embedded solitons exist on an infinite number of continuous curves 
in the $(v, \lambda)$ parameter plane. In other words, infinite continuous 
families of moving embedded solitons exist. These results greatly generalize the 
discrete set of embedded solitons reported in \cite{klauder, calvoakylas1}. 

First, we employ a transformation of variables 
so that one of the two free parameters ($v$ and $\lambda$) in Eq. (\ref{psi}) 
is removed. 
It may be tempting to just try rescaling $(\psi, \theta)$ variables, for
instance, defining $\psi/\sqrt{v}$ and $\sqrt{v}\:\theta$ 
(or $\psi/\sqrt{\lambda}$ and $\sqrt{\lambda}\:\theta$) as new variables. 
This does not work however: while $v$ (or $\lambda$) is normalized to 1, 
a new parameter appears in front of the $\psi_{\theta\theta\theta}$ term, 
thus no parameter reduction is achieved. A successful variable transformation
is the following: 
\begin{equation} \label{psitrans}
\psi(\theta)=\left(v+\frac{1}{3}\right)^{\frac{3}{4}}e^{-\frac{1}{3}i\theta}\Psi(\Theta), 
\end{equation}
\begin{equation}
\Theta=\left(v+\frac{1}{3}\right)^{\frac{1}{2}}\theta, 
\end{equation}
\begin{equation} \label{Lambda}
\Lambda=\left(v+\frac{1}{3}\right)^{-\frac{3}{2}}\left(\lambda+\frac{1}{3}v+\frac{2}{27}\right).
\end{equation}
Under this gauge transformation, Eq. (\ref{psi}) becomes
\begin{equation} \label{Psi}
-\Lambda \Psi +|\Psi|^2\Psi=i(\Psi_{\Theta\Theta\Theta}+\Psi_{\Theta}). 
\end{equation}
We see that only one parameter, $\Lambda$, remains now. In addition, 
the $\Psi_{\Theta\Theta}$ term has disappeared. 
Note that if $\psi(\theta)$ possesses the symmetry (\ref{symmetry}), 
so does $\Psi(\Theta)$.

\begin{figure}[h]
\begin{center}
\parbox[t]{12cm}{\postscript{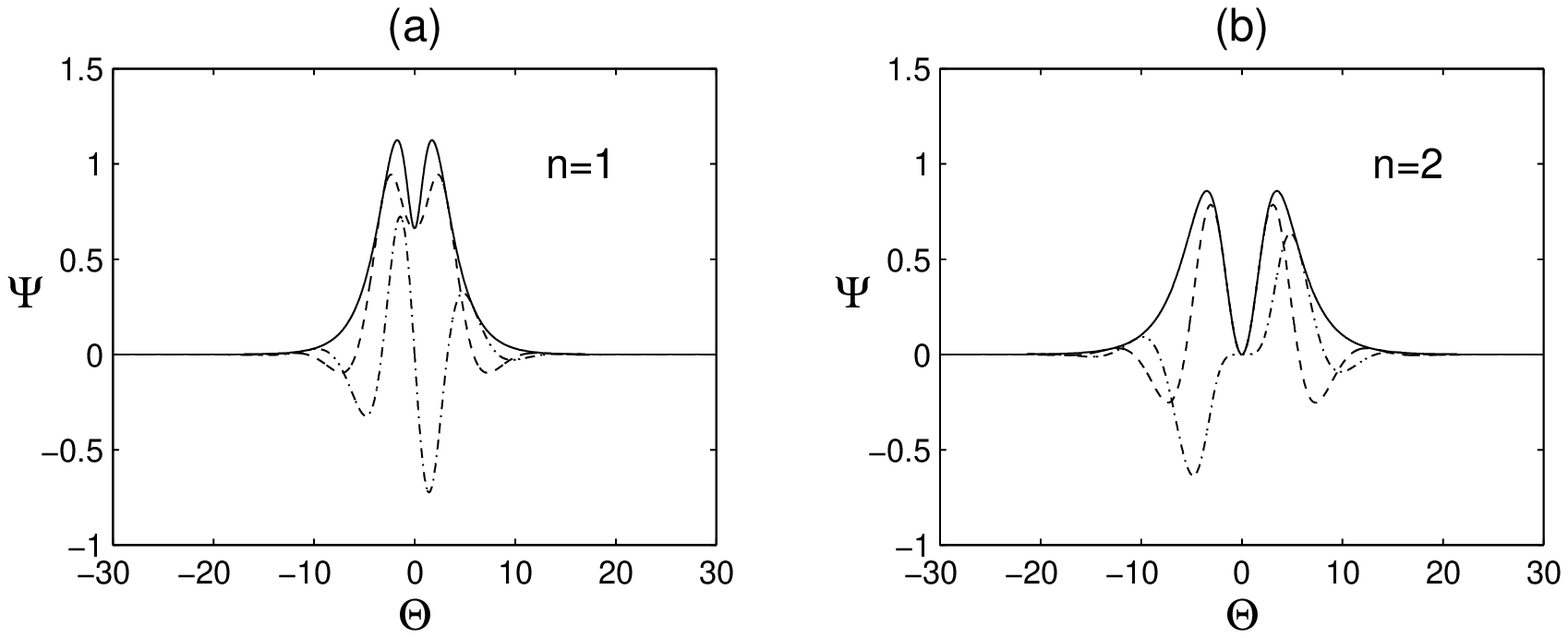}{1.0}}

\parbox[t]{12cm}{\postscript{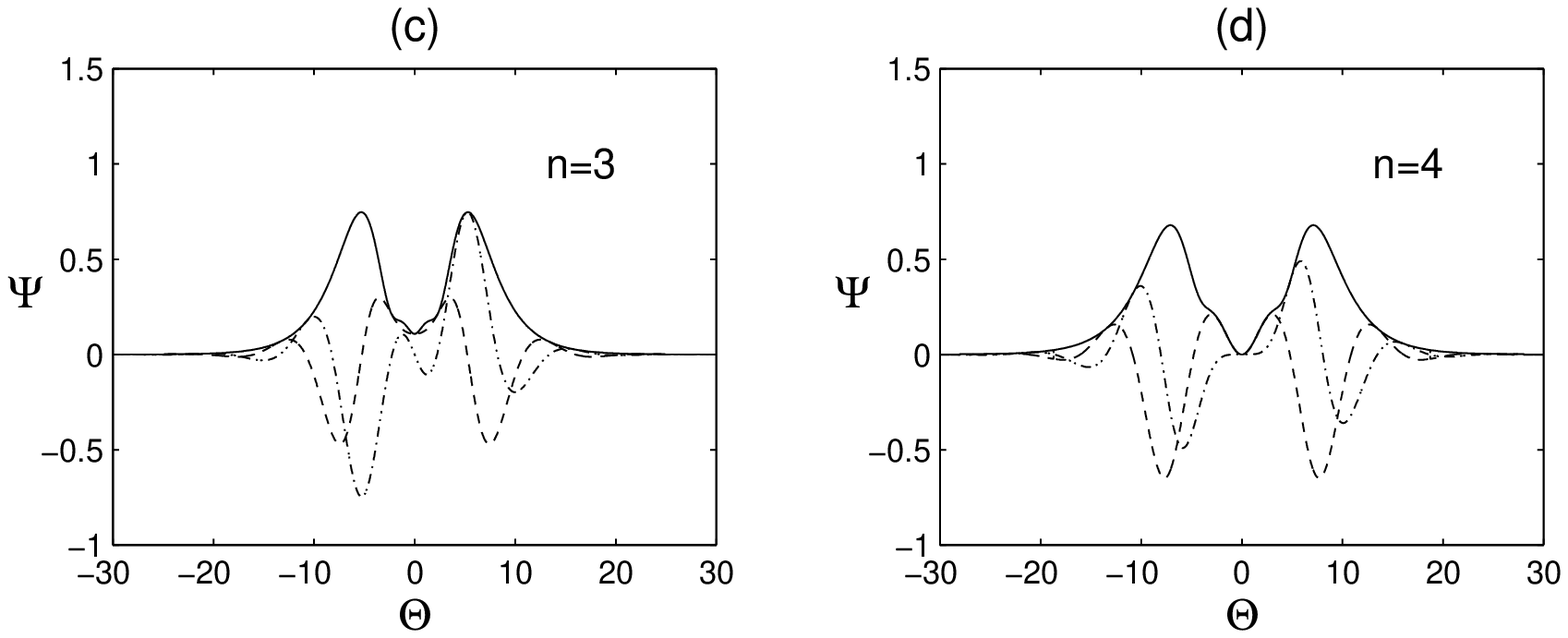}{1.0}}
\caption{The first four embedded solitons $\Psi(\Theta)$ in Eq. (\ref{Psi}). 
The corresponding $\Lambda_n$ values $(1\le n\le 4)$ are 0.8619, 
0.6959, 0.6316 and 0.5939 respectively. 
Solid lines: $|\Psi|$; dashed lines: Re($\Psi$); dash-dotted lines: Im($\Psi$). 
\label{Psifig} }
\end{center}  
\end{figure}

Eq. (\ref{Psi}) allows an infinite number of isolated, double-hump embedded solitons
at a discrete set of parameter values $\Lambda=\Lambda_n, \; 
(n=1, 2, \dots)$. These $\Lambda_n$ values can be inferred from previous
work on the related equation
\begin{equation} \label{u} 
u_{\theta\theta}-u+|u|^2u=i\epsilon (u_{\theta\theta\theta}-u_\theta), 
\end{equation}
where embedded solitons have been found at discrete $\epsilon$ values
$\epsilon_n \; (n=1, 2, \dots)$ \cite{klauder,calvoakylas1}. 
First, we introduce new variables $\hat{u}=\epsilon_nu$ and $\hat{\theta}=\theta/\epsilon_n$.
Then Eq. (\ref{u}) becomes the same as Eq. (\ref{psi}) with
$\lambda=\epsilon_n^2$ and $v=-\epsilon_n^2$. Substituting these $\lambda$ and $v$ values
into expression (\ref{Lambda}), we find that the equation (\ref{Psi}) for $\Psi$ admits 
embedded solitons at 
\begin{equation}
\Lambda_n=\left(\frac{1}{3}-\epsilon_n^2\right)^{-\frac{3}{2}}
\left(\frac{2}{3}\epsilon_n^2+\frac{2}{27}\right).
\end{equation}
As $n\to \infty$, $\epsilon_n \to 0$. In this limit, $\Lambda_n \to \Lambda_\infty \equiv
\frac{2\sqrt{3}}{9}\approx 0.3849$. 
The first few $\epsilon_n$ values were not explicitly given in \cite{klauder,calvoakylas1}. 
Thus, in order to get the corresponding $\Lambda_n$ values, we have used a shooting technique
on Eq. (\ref{Psi}) with boundary conditions similar to (\ref{psiinf}) and (\ref{psizero}). 
The first four $\Lambda_n$ are found to be 
$\Lambda_1=0.8619, \Lambda_2=0.6959, \Lambda_3=0.6316$, and $\Lambda_4=0.5939$. 
Embedded solitons at these four $\Lambda_n$ values are displayed in 
Fig. \ref{Psifig}. We see that at larger $n$, the soliton becomes lower, 
and the two humps tend to separate. 

\begin{figure}[h]
\begin{center}
\parbox[t]{12cm}{\postscript{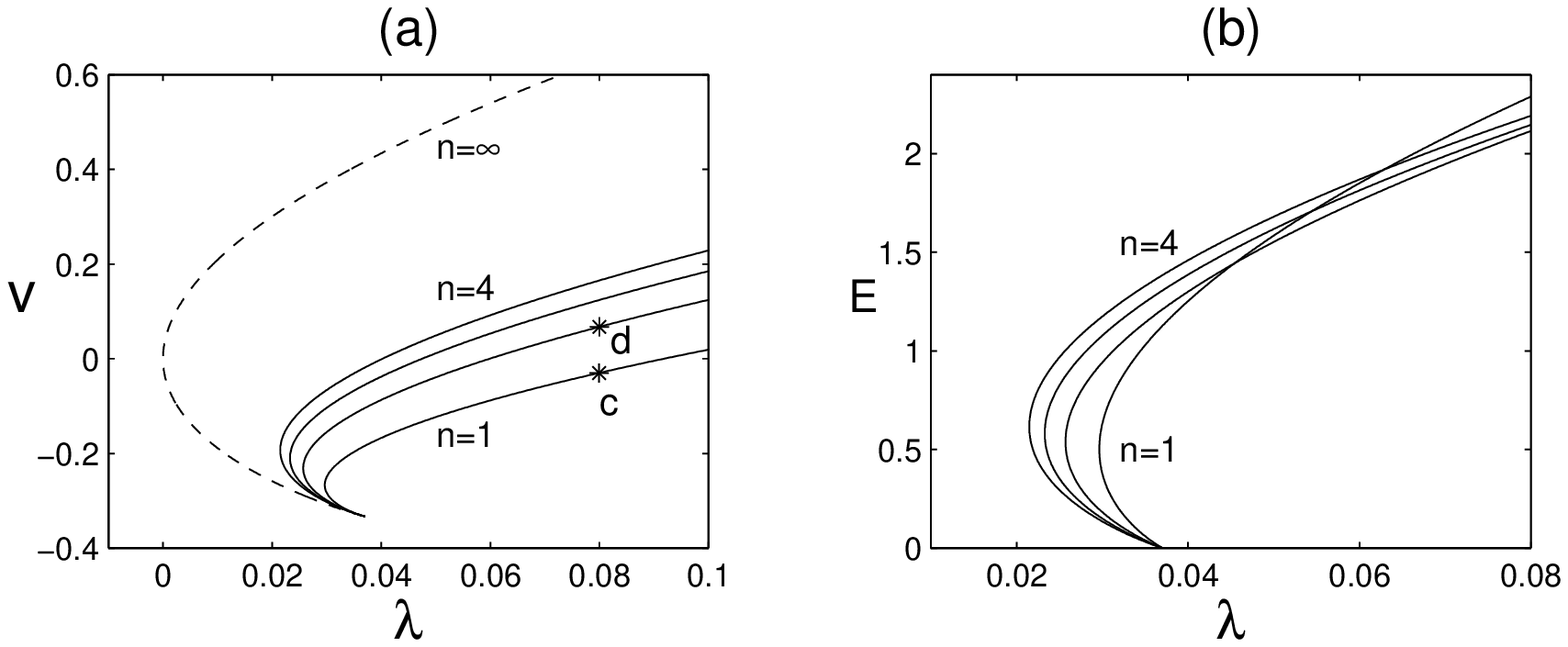}{1.0}}

\parbox[t]{12cm}{\postscript{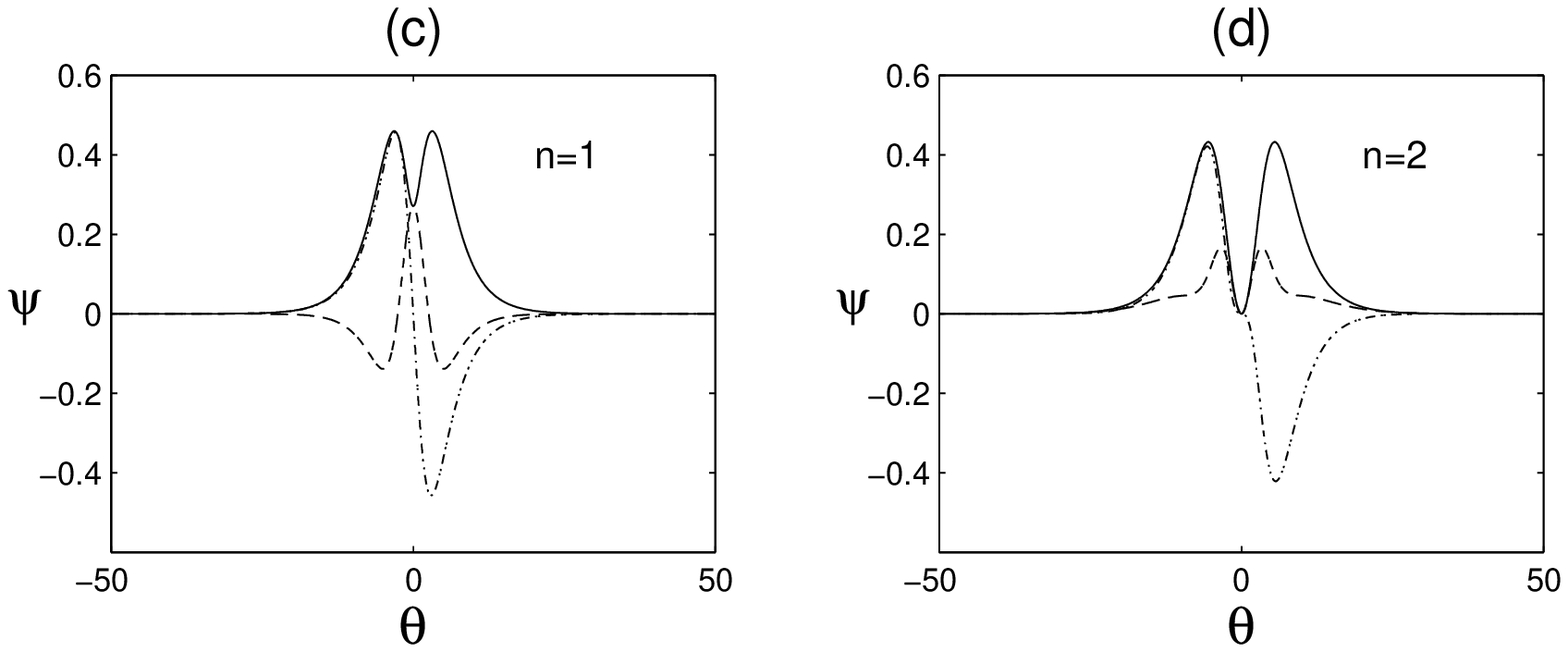}{1.0}}
\caption{(a, b) Families of curves in the parameter planes $(\lambda, v)$
and $(\lambda, E)$ where embedded solitons $\psi$ exist in the TNLS equation
(\ref{psi}). (c, d) Profiles of embedded solitons at velocity and frequency 
values marked as `c' and `d' in figure (a) respectively. 
Solid lines: $|\psi|$; dashed lines: Re($\psi$); dash-dotted lines: Im($\psi$). 
\label{curvefamily}}
\end{center}  
\end{figure}

Embedded solitons in equation (\ref{psi}) for $\psi$ can now be obtained from 
those in Eq. (\ref{Psi}) through the gauge transformation (\ref{psitrans}) -- (\ref{Lambda}). 
We find that embedded solitons $\psi$ exist on the following infinite number of
continuous curves in the $(v, \lambda)$ plane: 
\begin{equation}\label{lambdaform}
\lambda=\Lambda_n \left(v+\frac{1}{3}\right)^{\frac{3}{2}}-\frac{1}{3}v-\frac{2}{27}.
\end{equation}
The first four such curves ($1\le n\le 4$) as well as the limit curve ($n\to \infty$) 
are shown in Fig. \ref{curvefamily}(a). 
On each one of these curves, a continuous family
of embedded solitons $\psi$ exists. Note that the velocities of these families of
embedded solitons have a common lower bound, $v\ge -\frac{1}{3}$, 
but there are no upper bounds. Thus, for any velocity $v> -\frac{1}{3}$, 
a discrete infinite set of embedded solitons can be found. If $v< -\frac{1}{3}$, 
these solitons disappear. Frequency $\lambda$ also has a lower bound whose value
depends on the individual solution family. But $\lambda \ge 0$ holds for all families. 
On each solution curve, when $\lambda < \frac{1}{27}$ but above the lower $\lambda$ bound 
of that family, two embedded solitons with different velocities exist. 

The energy of an embedded soliton is an important quantity. Here we define
the energy as 
\begin{equation} \label{E}
E(\lambda, v)=\int_{-\infty}^\infty |\psi(\theta)|^2 d\theta. 
\end{equation}
Utilizing the gauge transformation (\ref{psitrans}), we easily find that 
\begin{equation}\label{EV}
E(\lambda, v)=\beta_n\left(v+\frac{1}{3}\right), 
\end{equation}
where $\beta_n=\int_{-\infty}^\infty|\Psi(\Theta)|^2d\Theta$. 
The first four $\beta_n$ values $(1\le n\le 4)$ are 
7.5434, 5.2786, 4.6916 and 4.3951 respectively. 
Formula (\ref{EV}) indicates that the energy of an embedded soliton
increases linearly with velocity. 
Eliminating $v$ from Eqs. (\ref{lambdaform}) and (\ref{EV}), we find that the energy $E$ 
is related to the frequency $\lambda$ as 
\begin{equation}
\lambda=\Lambda_n \left(\frac{E}{\beta_n}\right)^{\frac{3}{2}}-
\frac{E}{3\beta_n}+\frac{1}{27}.
\end{equation}
The first four $(\lambda, E)$ curves are displayed
in Fig. \ref{curvefamily}(b). Note that on each energy curve, when 
$\lambda < \frac{1}{27}$, two embedded solitons with different energies exist. 

The profiles of embedded solitons $\psi(\theta)$ in the above solution families 
can be readily deduced from the embedded solitons $\Psi(\Theta)$ through the
gauge transformation (\ref{psitrans}) -- (\ref{Lambda}). 
To illustrate typical embedded solitons $\psi$ of Eq. (\ref{psi}), 
we select two points in Fig. \ref{curvefamily}(a)
marked as `c' and `d', which belong to the first and second solution families
respectively. The coordinates of these two points are approximately 
$(\lambda, v)=(0.08, -0.03)$ and $(0.08, 0.0674)$. 
Embedded solitons at these two points are shown in 
Fig. \ref{curvefamily}(c, d). We see that their amplitude profiles ($|\psi|$)
are similar to those in Fig. \ref{Psifig}(a, b) save for horizontal and vertical 
rescalings, while their phase distributions are different. 
The reason is apparently due to the gauge transformation 
(\ref{psitrans}) -- (\ref{Lambda}).

\begin{figure}[h]
\begin{center}
\parbox[t]{12cm}{\postscript{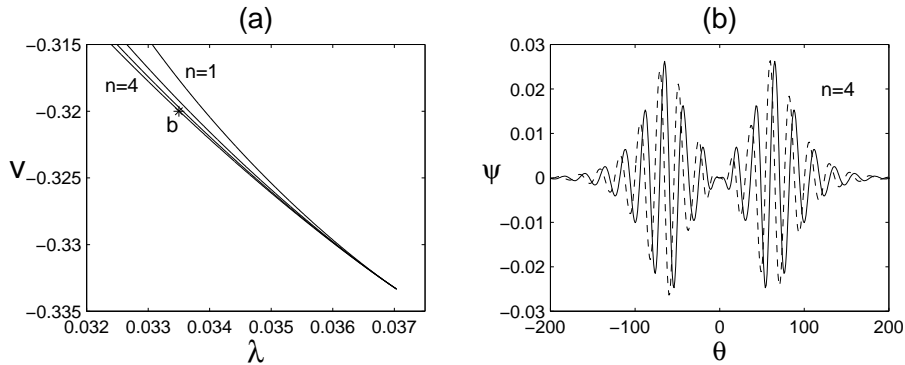}{1.0}}
\caption{Embedded solitons near the bifurcation point ($\lambda_c, v_c)=(1/27, -1/3)$. 
(a) enlargement of $(\lambda, v)$ curves for embedded solitons near this point; 
(b) an embedded soliton of the fourth family at the point marked
   `b' in figure (a). Solid line: Re($\psi$); dashed line: Im($\psi$). 
\label{familyedge}}
\end{center}  
\end{figure}

It is interesting to note that all solution families in Fig. 
\ref{curvefamily}(a) emanate from the single point 
$(\lambda_c, v_c)=(\frac{1}{27}, -\frac{1}{3})$ which, 
according to (\ref{psitrans}), corresponds to the linear limit
($|\psi|\to 0$) of those soliton solutions. The significance of this
critical point may also be seen by considering infinitesimal normal-mode
disturbances $e^{i\kappa\theta}$ and examining the linear spectrum of 
Eq. (\ref{psi}): 
\begin{equation} \label{dispersion}
F(\kappa; \lambda, v)\equiv \kappa^3+\kappa^2-v\kappa+\lambda=0. 
\end{equation}
Clearly, (\ref{dispersion}) has either three real or one real and a pair of
complex conjugate roots. Based on prior experience 
\cite{akylas93,grimshawmalomed,YangAkylas}, one then would expect 
small-amplitude solitary waves, in the form of wave packets, to 
bifurcate from infinitesimal sinusoidal disturbances having
(real) wavenumber $\kappa_c$ that corresponds to a triple root of the 
linear spectrum: 
\begin{equation} \label{triple}
F(\kappa_c; \lambda_c, v_c)=0, \hspace{0.3cm}
F'(\kappa_c; \lambda_c, v_c)=0, \hspace{0.3cm}
F''(\kappa_c; \lambda_c, v_c)=0. 
\end{equation}
The critical point $(\lambda_c, v_c)$ in Fig. \ref{curvefamily}(a) is
consistent with these conditions, and the value of the critical 
wavenumber is found to be $\kappa_c=-\frac{1}{3}$. The second of conditions
(\ref{triple}), in particular, implies that the ``phase speed'' $v$ of
linear sinusoidal waves is stationary at critical conditions and hence
is equal to the group velocity there. 
(Note: in this interpretation of phase speed and group speed, 
the frequency shift $\lambda$ introduced in Eq. (\ref{phiform}) 
is treated as a free parameter.)
Accordingly, small-amplitude solitary wavepackets close to the bifurcation
point may also be interpreted as envelope solitons with stationary 
crests \cite{akylas93,grimshawmalomed,YangAkylas}.
Fig. \ref{familyedge} displays one such embedded-soliton wavepacket
which belongs to the fourth family. 
One difference between wave packets here and those 
in \cite{akylas93,grimshawmalomed,YangAkylas}, however, is that 
the present wave packets are embedded solitons, while
those in \cite{akylas93,grimshawmalomed,YangAkylas} are not.

We remark here that the infinite 
families of embedded solitons displayed in Fig. \ref{curvefamily}
are not the only possible embedded solitons $\psi$ of Eq. (\ref{psi}). Equivalently, 
the infinite discrete set of embedded solitons, the first four of which
are displayed in Fig. \ref{Psifig}, are not the only possible 
embedded solitons $\Psi$ of Eq. (\ref{Psi}). The solitons
we have studied above have two major humps in their amplitude profiles. 
Calvo and Akylas \cite{calvoakylas1} have shown that embedded solitons with
three or more major humps exist as well. In this paper, we shall not
discuss those more complicated solitons.

\section{Linear and nonlinear stability of embedded solitons}
We now turn to the stability properties of the families of embedded solitons
found in the last section. This problem has been considered 
in \cite{klauder,calvoakylas2} for certain isolated embedded solitons. 
It was suggested that they are linearly weakly unstable, but
nonlinearity has a stabilizing effect, permitting those solitons
to propagate for long time without breakup \cite{calvoakylas2}. 
On the other hand, we have seen in the above section 
that when $\lambda < \frac{1}{27}$, there exist two branches of embedded solitons 
in the same family having different energy [see
Fig. \ref{curvefamily}(b)]. In such a case, 
it is typically expected from saddle-node bifurcations in dynamical systems
that one branch of solutions is stable, while the other branch is
unstable \cite{peli}. 
If this holds also for the TNLS equation, then one would expect
different stability properties for the two branches of embedded solitons
in the same solution family. Previous work also shows that while
single-hump (fundamental) solitons are often linearly stable, 
multi-hump solitons (whether embedded or not) are often linearly
unstable \cite{YangPRL99,YangChampneys}. 
This fact suggests that embedded solitons in the TNLS equation, or at least
higher families ($n\ge 2$) of such solitons, might be linearly unstable. 
If embedded solitons are linearly stable, nonlinearly they could be semi-stable, 
i.e., whether they persist or break up depends on the
type of initial perturbations imposed. 
This semi-stability property has been established rigorously 
for isolated embedded solitons in Hamiltonian systems
\cite{YangPRL99,tan,PeliYang,ChampneysYang,YangChampneys}. 
For the TNLS equation, embedded solitons exist as continuous families. 
Thus, when they are perturbed, they may be able to shed some energy via radiation
and tend to nearby embedded solitons in the same solution family. 
If this happens, these embedded solitons could be nonlinearly stable. 

In this section, we will establish, however, that the stability properties of
embedded solitons in the TNLS equation (\ref{phi}) do not follow 
the above common scenarios. 
In particular, we will show that for the TNLS equation, 
(i) all embedded solitons in the same family have identical linear and
nonlinear stability properties; (ii) all families of embedded solitons are linearly stable;  
(iii) embedded solitons are nonlinearly semi-stable at least for the first and second families. 
The first result can be easily seen from the fact that 
all embedded solitons in the TNLS equation are related to each other
by the gauge transformation (\ref{psitrans}) -- (\ref{Lambda}), thus
they must be either all stable or all unstable. 
This contrasts other physical systems where different branches 
in the same solution family have different stability properties \cite{peli}. 
The second and third results will be established in the following two subsections. 

\subsection{Linear stability of embedded solitons}
To study the linear stability of embedded solitons (\ref{phiform}) in the TNLS 
equation (\ref{phi}), we write
\begin{equation}
\phi(x, t)=e^{i\lambda t}\left\{\psi(\theta)+\tilde{\phi}(\theta, t)\right\}, 
\end{equation}
where $\tilde{\phi}$ is an infinitesimal perturbation, and $\theta$ is defined in 
(\ref{theta}). The linearized equation for $\tilde{\phi}$ is
\begin{equation} \label{phitilde}
i\tilde{\phi}_t-\lambda\tilde{\phi}-iv\tilde{\phi}_\theta+\tilde{\phi}_{xx}+
2|\psi|^2\tilde{\phi}+\psi^2\tilde{\phi}^*-i\tilde{\phi}_{\theta\theta\theta}=0. 
\end{equation}
To determine the linear stability of these solitons, we numerically simulate the above 
linearized equation to see if its solution $\tilde{\phi}$ has exponentially 
growing modes or not. Since a whole family of embedded solitons 
has the same stability behavior, we only need to pick one embedded soliton
from each family and test its linear stability. The numerical scheme we use
is the pseudo-spectral method (FFT) along the $x$-direction, and the
fourth-order Runge--Kutta method in $t$. 
For simplicity, we choose a Gaussian initial condition
\begin{equation} \label{linearic}
\tilde{\phi}(\theta, 0)=(1+i)e^{-\frac{1}{2}\theta^2}. 
\end{equation}
Other initial conditions have been used as well, and the results are
qualitatively the same. 

\begin{figure}[h]
\begin{center}
\parbox[t]{12cm}{\postscript{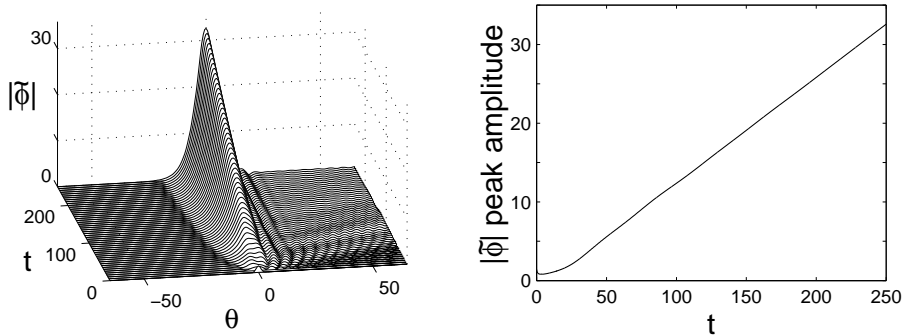}{1.0}}

\caption{Evolution of infinitesimal perturbations to the embedded soliton of the 
first family, shown in Fig. \ref{curvefamily}(c).  
\label{linearevolution1} }
\end{center}  
\end{figure}

In the first family of embedded solitons ($n=1$, see Fig. \ref{curvefamily}), 
we choose the soliton as displayed in Fig. \ref{curvefamily}(c), whose
frequency and velocity values are $(\lambda, v)=(0.08, -0.03)$. 
For this soliton, the evolution of the disturbance $\tilde{\phi}$ is shown 
in Fig. \ref{linearevolution1}. 
Note that Fig. \ref{linearevolution1}(a) 
is qualitatively the same as Fig. 3 in Ref. \cite{calvoakylas2}. 
We see that the disturbance grows. In \cite{calvoakylas2}, this was 
interpreted as weak exponential growth. 
However, Fig. \ref{linearevolution1}(b)
reveals that the disturbance grows only linearly. This linear growth just
corresponds to an adjustment of this soliton's frequency and velocity values,
and it is not a sign of linear instability. Thus, this embedded soliton, 
or equivalently the first family of embedded solitons, is linearly stable. 

It is not difficult to determine the origin of this linear growth in the
disturbance $\tilde{\phi}$. In fact, the linearly growing mode
\begin{equation}
\tilde{\phi}(\theta, t)=\alpha_1 \left[i\psi t+\frac{\partial \psi}{\partial 
\lambda}\right]_{(\lambda_0, v_0)}+\alpha_2 \left[\frac{\partial \psi}{\partial \theta}t
-\frac{\partial \psi}{\partial
v}\right]_{(\lambda_0, v_0)}
\end{equation}
is a particular solution of the linearized equation
 (\ref{phitilde}). Here $\alpha_1$ and $\alpha_2$ are
arbitrary real constants, $\psi=\psi(\theta, \lambda, v)$ is the solution of Eq. (\ref{psi})
for general $(\lambda, v)$ values (which is generally nonlocal), and
$(\lambda_0, v_0)$ are the embedded soliton's frequency and velocity parameters. 
A general initial perturbation $\tilde{\phi}(\theta, 0)$ excites this mode, 
thus the disturbance grows linearly in time. But this mode does not imply 
exponential instability. An analogous situation is the NLS soliton under perturbations
(see \cite{Kaup90,Yang97}). 

\begin{figure}[h]
\begin{center}
\parbox[t]{12cm}{\postscript{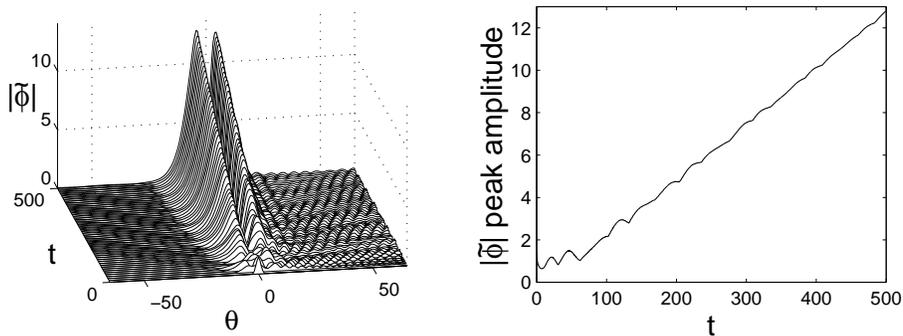}{1.0}}

\caption{Evolution of infinitesimal perturbations to the embedded soliton of the 
second family, shown in Fig. \ref{curvefamily}(d).  
\label{linearevolution2} }
\end{center}  
\end{figure}

Are other families of embedded solitons with $n\ge 2$ linearly stable?
To answer this question, we have repeated the above numerical simulation
for embedded solitons in the higher families, and found that they are 
{\em all} linearly stable. To illustrate, we consider the second family, and
pick the embedded soliton as shown in Fig. \ref{curvefamily}(d), 
where the frequency and velocity parameters are $(\lambda, v)=(0.08, 0.0674)$. 
With the same Gaussian initial condition (\ref{linearic}), evolution of
the disturbance $\tilde{\phi}$ is displayed in Fig. \ref{linearevolution2}. 
We see that, just like the first family, the disturbance grows linearly, 
which implies that the second solution family is linearly stable as well. 

\subsection{Nonlinear semi-stability of embedded solitons}
As we have shown above, 
embedded solitons in the TNLS equation (\ref{phi}) exist as continuous
families, and they are all linearly stable. 
Are they also nonlinearly stable to small perturbations?
If embedded solitons are isolated in a Hamiltonian system, these solitons 
are generally semi-stable \cite{YangPRL99,tan,PeliYang,ChampneysYang,YangChampneys}. 
The reason can be understood heuristically as follows \cite{YangPRL99}. 
When an isolated embedded soliton is perturbed, it will tend to an adjacent state 
which is not locally confined, but features tails of non-zero amplitude. 
However, it would require infinite energy to create a non-vanishing tail. 
Because the energy is conserved in a Hamiltonian system, and 
the energy $E_{0}$ of the unperturbed embedded soliton is
finite, there are two possibilities: first, if the initial perturbed state
has a total energy $E<E_{0}$, the energy lost in an attempt to generate the
infinite tail will drive the soliton farther away from the initial state. As
a result, the perturbed embedded soliton can be expected to eventually decay into
radiation. On the other hand, if the initial perturbed state has energy $E>E_{0}$, we
may expect the energy lost in generating the tail to drag the pulse back
toward the unperturbed embedded soliton. Thus, we can anticipate that the 
embedded soliton is subject to a non-exponential {\it one-sided} instability
which is the so-called semi-stability. 

However, embedded solitons in the TNLS equation (\ref{phi}) are continuous
rather than isolated. These solitons have a free parameter which is their velocity. 
In addition, their energy depends linearly on their velocity [see Eq. (\ref{EV})], 
thus their energy can acquire an arbitrary value. Because of this, 
the above argument for semi-stability no longer holds for embedded solitons in the TNLS equation. 
When an embedded soliton in the TNLS equation is perturbed, in principle, it
could simply emit some radiation and adjust its shape to a nearby embedded soliton
in the same solution family, as an NLS soliton does under perturbations. 
This prospect originally led us to speculate
that embedded solitons in the TNLS equation may actually be nonlinearly stable.
Unfortunately, this speculation turns out to be incorrect. Our numerical simulations
show that
these solitons are still semi-stable, just like isolated embedded solitons in other
physical systems \cite{YangPRL99,YangStudies,tan,PeliYang}. In other words, 
the freedom of arbitrary velocities of embedded solitons in the TNLS equation
is not sufficient to stabilize these solitons. 

To explore the nonlinear stability of embedded solitons in the TNLS 
equation, we numerically simulate this equation starting with an embedded soliton
under perturbations. Since a whole family of embedded solitons have the same
stability properties, it is sufficient to pick one soliton from each family and
test its stability. In numerical simulations, we adopt the coordinates which move
at the speed $v$ of the embedded soliton. Then the TNLS equation (\ref{phi}) becomes  
\begin{equation}
i\phi_t-iv\phi_\theta+\phi_{\theta\theta}+|\phi|^2\phi=i\phi_{\theta\theta\theta},
\end{equation}
where $\theta$ has been defined in Eq. (\ref{theta}). In these coordinates, 
an embedded soliton is given by (\ref{phiform}). 
Consistent with the initial perturbations which we have used in other
wave systems which admit embedded solitons \cite{YangPRL99,tan,PeliYang,YangChampneys}, 
we use the initial condition 
\begin{equation} \label{nonlinearic}
\phi(\theta, 0)=(1+\alpha) \psi(\theta),
\end{equation}
where $\alpha$ is a small real constant. 
That is, our initial perturbed state is the original soliton amplified by
a factor $1+\alpha$. In this case, the energy of the initial state
is $(1+\alpha)^2$ multiplied by the energy of the unperturbed soliton [see Eq. (\ref{E})]. 
When $\alpha>0$, the perturbed state has higher energy than the unperturbed
soliton, while when $\alpha<0$, the perturbed state has lower energy than 
the unperturbed soliton. In previous studies on isolated embedded solitons, 
we have termed the former perturbations as ``energy increasing'', 
and the latter perturbations as ``energy decreasing'' \cite{YangPRL99,tan,PeliYang}. 
However, in the present case, we will refrain from using such terms. The reason is
that embedded solitons here are continuous, so is their energy. 
Thus it may be ambiguous or even confusing to say 
``energy-increasing or -decreasing perturbations''. 
Other initial perturbations different from (\ref{nonlinearic}) 
can also be taken, but they are not expected to change the qualitative conclusions. 

Because of the gauge transformation (\ref{psitrans}) -- (\ref{Lambda}), 
it makes no difference which embedded soliton $\psi(\theta)$ in a solution family 
we use in our simulations. In other words, for a fixed value of $\alpha$, 
whether the perturbed state (\ref{nonlinearic}) eventually
breaks up or persists is independent of the choice of the embedded soliton in a 
solution family. 

\begin{figure}[h]
\begin{center}
\parbox[t]{5cm}{\postscript{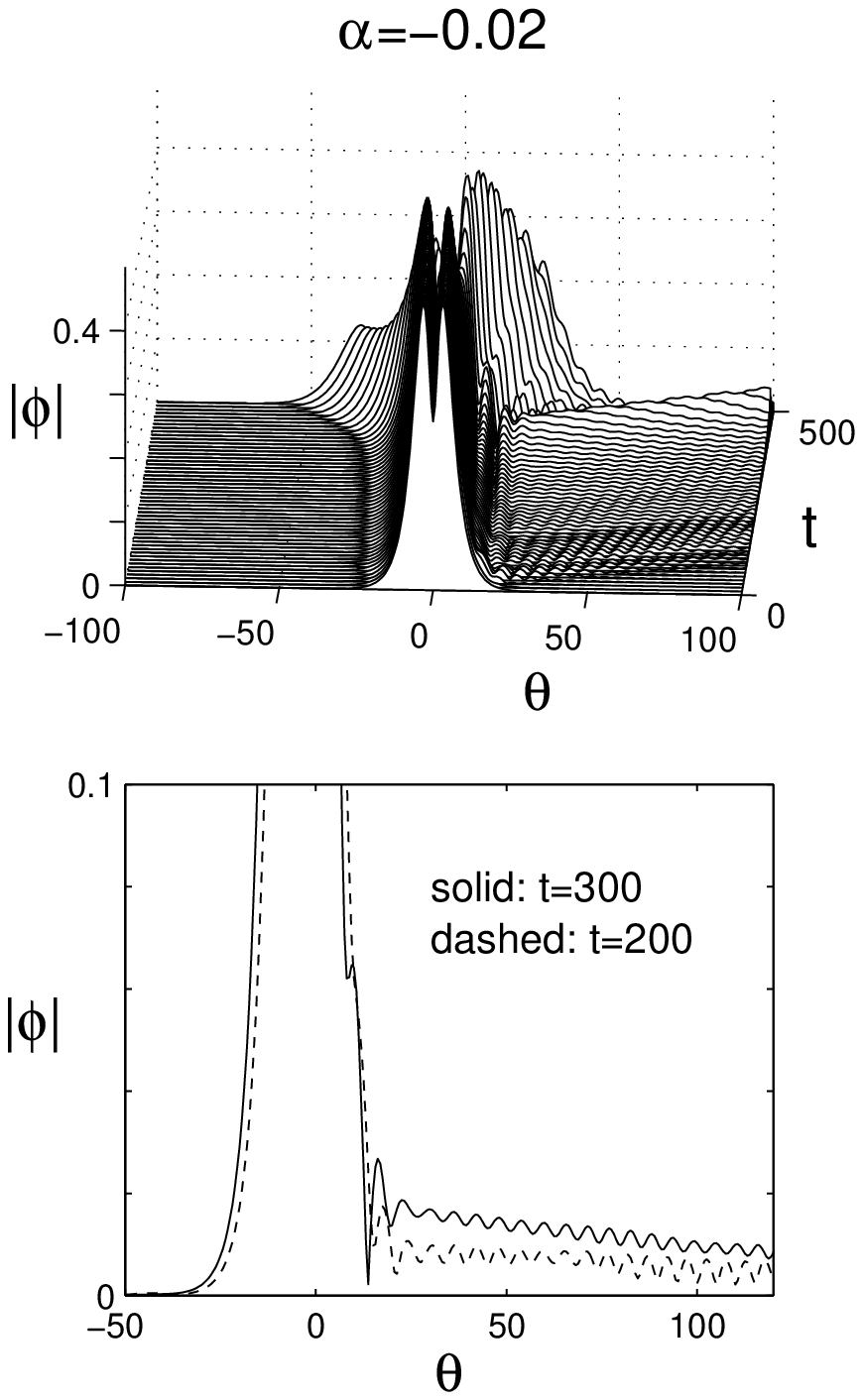}{1.0}} \hspace{0.5cm}
\parbox[t]{5cm}{\postscript{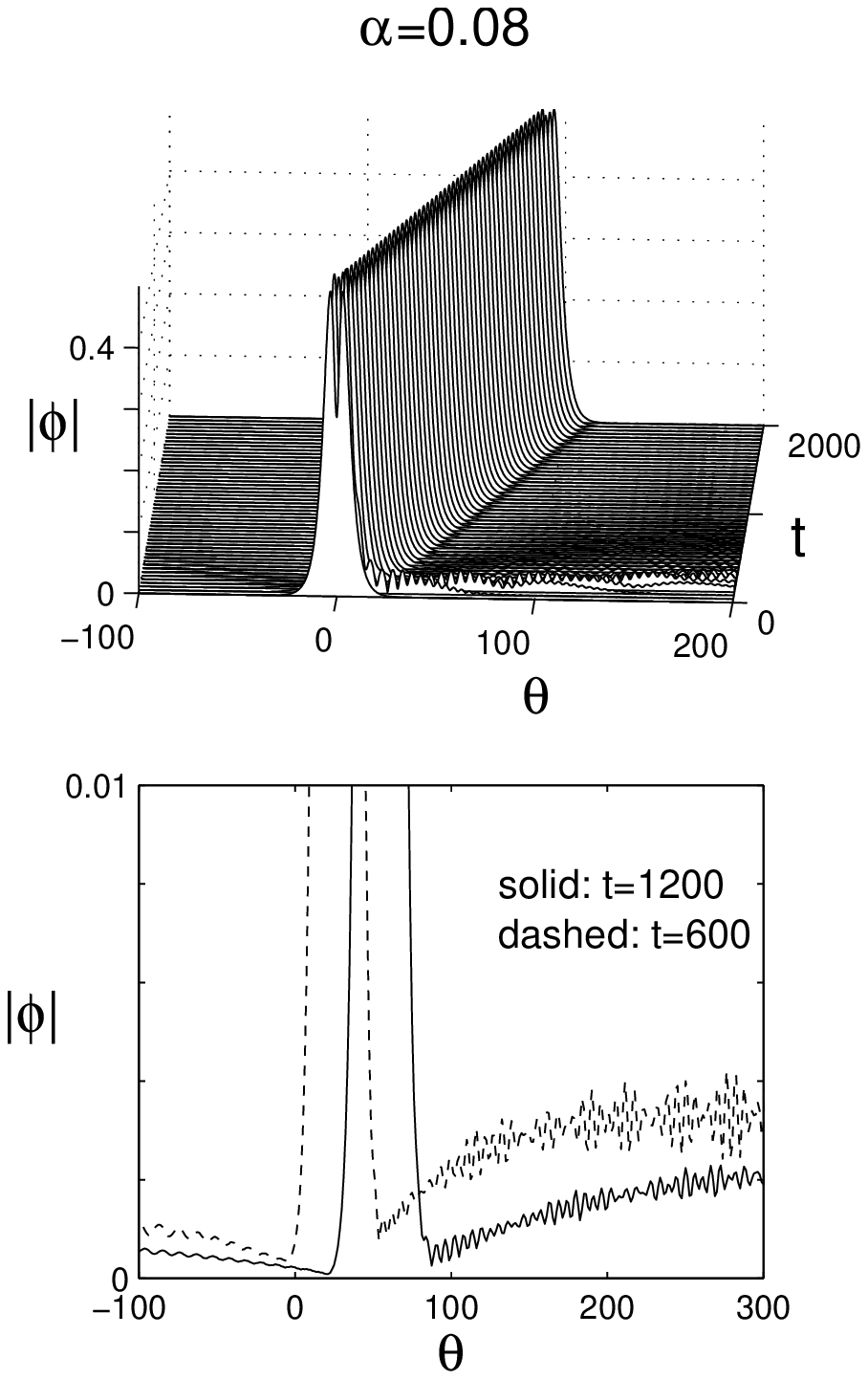}{1.0}} \hspace{0.5cm}
\parbox[t]{5cm}{\postscript{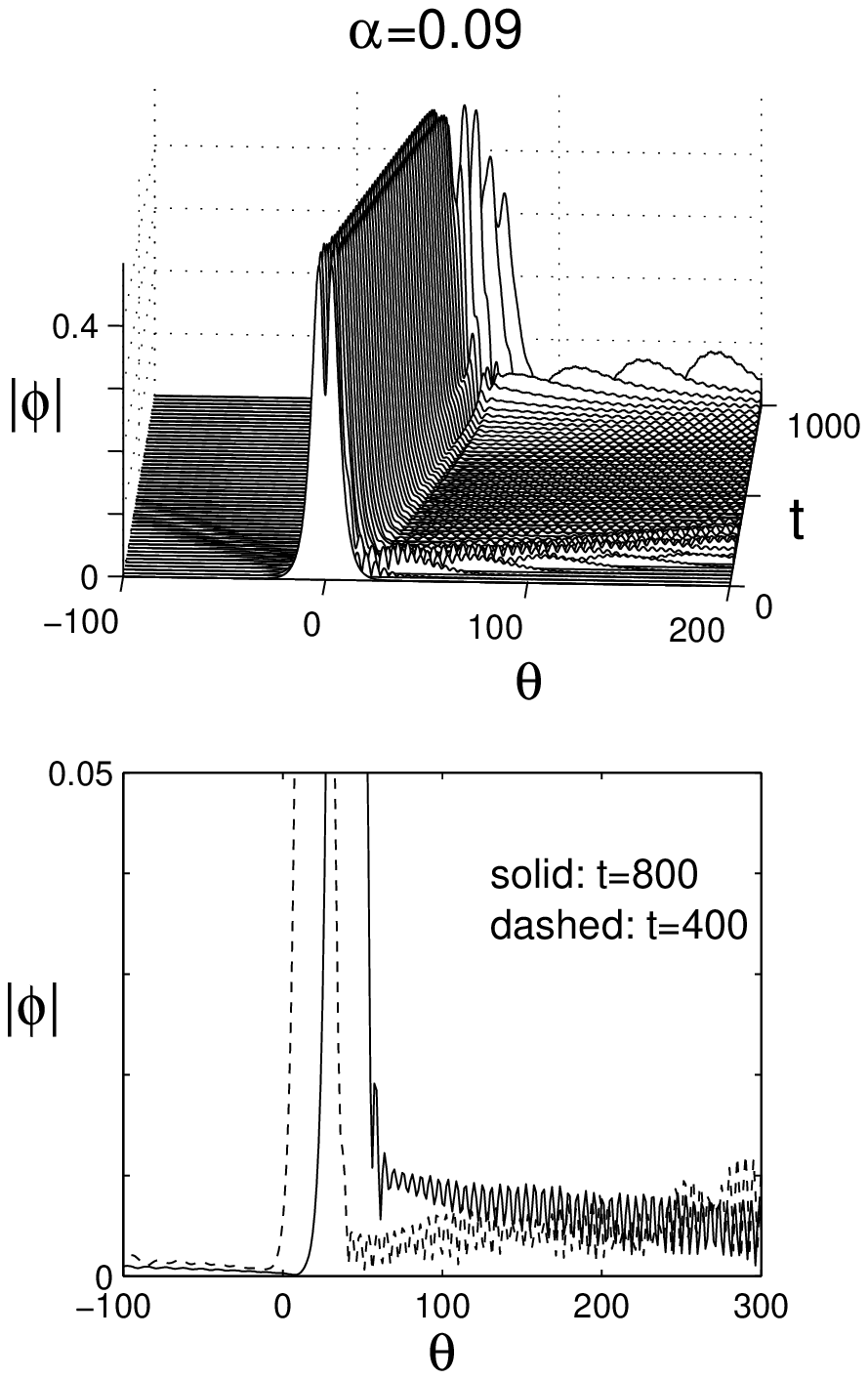}{1.0}}
\caption{Evolution of the embedded soliton of the first family, 
shown in Fig. \ref{curvefamily}(c), 
under perturbations (\ref{nonlinearic}) for various values of $\alpha$. 
First column: $\alpha=-0.02$;  second column: $\alpha=0.08$; third column: $\alpha=0.09$.  
\label{evolutionn1} }
\end{center}  
\end{figure}

We first consider the first family of embedded solitons. 
Extensive numerical simulations have shown that this family of 
embedded solitons persist when $0\le \alpha \:^<_\sim \: 0.081$, while they break up otherwise. 
The simulation results for three values of $\alpha$, $-0.02$, 0.08 and 0.09, are plotted 
in Fig. \ref{evolutionn1}, where the particular soliton used is as 
displayed in Fig. \ref{curvefamily}(c) with $(\lambda, v)=(0.08, -0.03)$. 
We see that for $\alpha=-0.02$, the embedded soliton sheds
continuous-wave radiation whose amplitude steadily increases over time. As a result, the soliton
breaks up.  This behavior is typical of 
isolated embedded solitons under energy-decreasing perturbations \cite{YangPRL99,tan,PeliYang}. 
For $\alpha=0.08$, the soliton also sheds continuous-wave radiation, but the radiation
tail decreases over time. Meanwhile, the central pulse adjusts itself and approaches a 
nearby embedded soliton with a higher velocity. As a result, the soliton persists under this 
perturbation. This behavior is somewhat similar to 
isolated embedded solitons under energy-increasing perturbations \cite{YangPRL99,tan,PeliYang}. 
However, the new feature here is that the final embedded soliton is different from the
unperturbed soliton, although the final soliton clearly still belongs to the first family. 
What happens here 
is that part of the increased energy in the initial perturbed state is absorbed by the 
original soliton and changes it to a nearby soliton with higher energy (velocity),    
while the rest of the increased energy radiates away. 
For $\alpha=0.09$, however, the situation is different. Initially, the perturbed state
appears to adjust itself toward a nearby soliton with higher energy (velocity). But later on,
the tail radiation starts to increase, which eventually breaks up the soliton. 
We have checked all these simulations with higher accuracy and over longer periods of time, and
the results remain the same. 
It is noted that our numerical results above are consistent with previous 
numerical simulations by Calvo and Akylas \cite{calvoakylas2}. 

The above numerical results indicate that the first family of embedded solitons are
semi-stable: whether the soliton persists or breaks up depends on 
the initial perturbation. Compared to the dynamics of isolated embedded solitons, 
a new feature we find here is that embedded solitons are unstable not only for $\alpha < 0$, 
but also for $\alpha$ above a certain threshold value (which is 0.081 here). 
At the present time, the reason for this new behavior is still not clear. 
It could be because when $\alpha > 0.081$, the perturbation becomes too strong
for this double-humped embedded soliton. As we know, any soliton can be broken up
with a strong enough perturbation. But the fact that the perturbed state initially
does adjust itself toward a nearby soliton makes us suspect that the reason for its
eventual breakup may be elsewhere [see Fig. \ref{evolutionn1}(c)]. 
A full explanation for this new behavior may be obtained from a detailed internal-perturbation
calculation as has been done for isolated embedded solitons in several other physical systems
\cite{tan,PeliYang}. But this remains to be seen.   

\begin{figure}[h]
\begin{center}
\parbox[t]{5cm}{\postscript{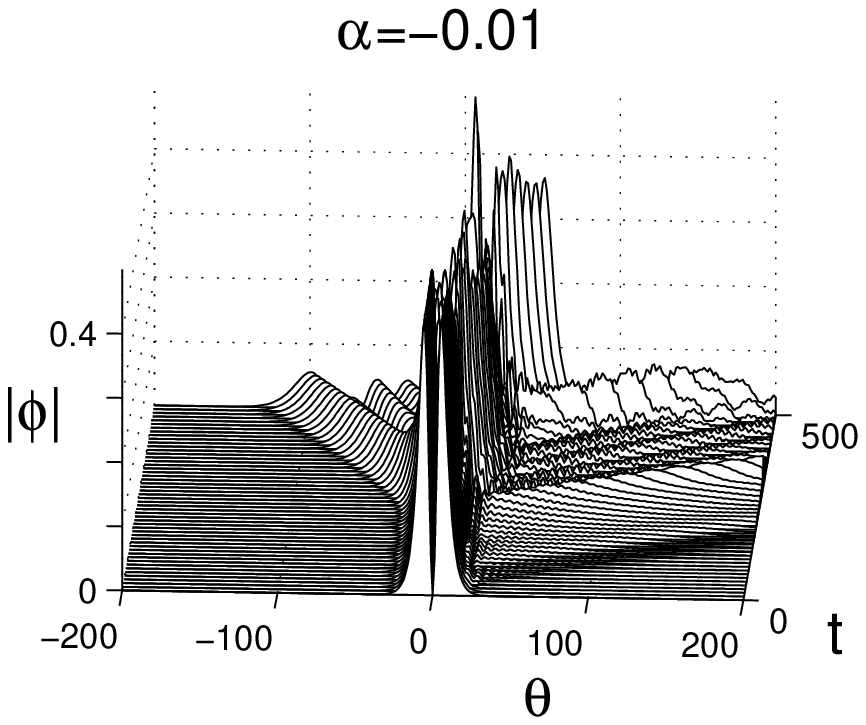}{1.0}} \hspace{0.5cm}
\parbox[t]{5cm}{\postscript{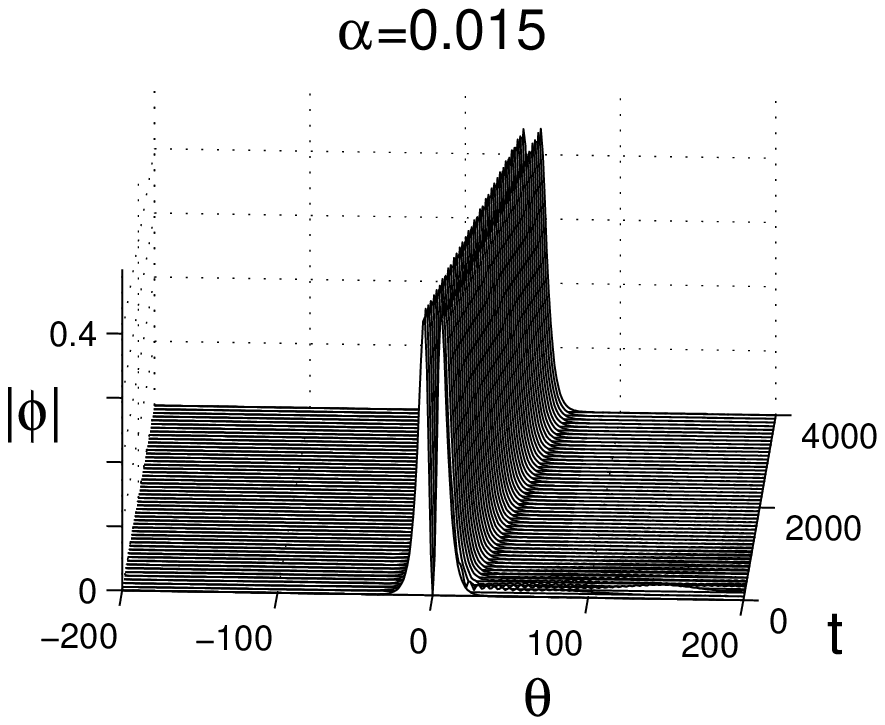}{1.0}} \hspace{0.5cm}
\parbox[t]{5cm}{\postscript{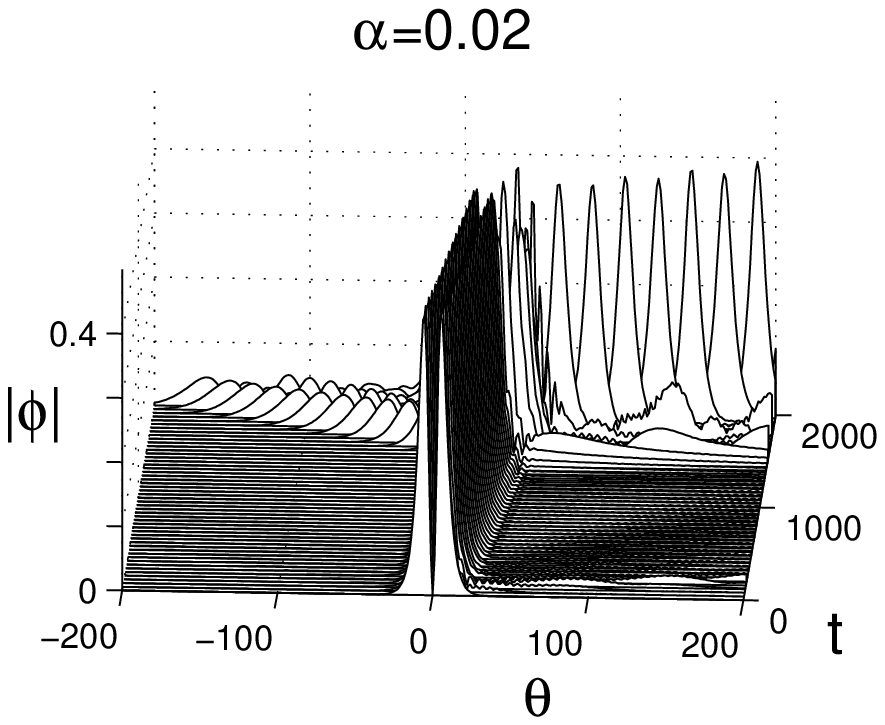}{1.0}}
\caption{Evolution of the embedded soliton of the second family,
shown in Fig. \ref{curvefamily}(d),
under perturbations (\ref{nonlinearic}) for various values of $\alpha$.
\label{evolutionn2} }
\end{center}  
\end{figure}

Are higher families of embedded solitons also semi-stable? To explore this question, 
we have repeated the above simulations for an embedded soliton in the second family. 
The initial perturbation remains the same as in (\ref{nonlinearic}). In this case, 
we have found that the second family of embedded solitons is stable only when 
$0\le \alpha \:^<_\sim \: 0.018$, and
breaks up otherwise. Three simulation results with $\alpha=-0.01, 0.015$ and 0.02
are plotted in Fig. \ref{evolutionn2}, where the particular embedded soliton used is the one
shown in Fig. \ref{curvefamily}(d) with $(\lambda, v)=(0.08, 0.0674)$. 
These results are qualitatively the same as
for the first family, but the window of $\alpha$ for soliton stability is much narrower. 
In other words, embedded solitons of the second family are more prone to breakup under
perturbations. For the third family of embedded solitons, our numerical simulations did not
find a window of $\alpha$ for soliton persistence. This may be because that window
is too small and we did not detect it. It is also possible that 
such a window actually disappears for the third (and higher)
families of embedded solitons. This issue is not pursued further in this paper.  

\section{Discussion}
In this article, we have studied embedded solitons and their stability in the
third-order NLS equation (\ref{phi}). We have discovered an infinite number of 
continuous families of embedded solitons parameterized by their velocities, 
or equivalently by their energies. 
We have further shown that these families of embedded solitons are all linearly
stable. But nonlinearly they are still semi-stable, just like isolated 
embedded solitons in other physical systems \cite{YangPRL99,tan,PeliYang}.

In the theory of embedded solitons, the following question still remains open: 
are nonlinearly stable embedded solitons in Hamiltonian systems possible or not?
This question is quite important for physical applications. 
Previous work has made it clear that 
a necessary condition for 
such embedded solitons to be possible is that they exist 
as continuous families, not as isolated solutions \cite{YangPRL99,tan,PeliYang}. 
However, our results in this paper indicate that this condition is apparently
not sufficient. Whether other physical systems support nonlinearly 
stable embedded solitons or not needs further investigation.

\section*{\hspace{0.1cm} Acknowledgments}
The work of J.Y. was partially supported by the National Science Foundation
under grant DMS-9971712, and by a NASA EPSCoR minigrant. The work of T.R.A. 
was partially supported by the Air Force Office of Scientific Research, 
Air Force Materials Command, USAF, under grant F49620-01-1-001, and by
the National Science Foundation under grant DMS-0072145.

\end{document}